\documentclass[sigconf,screen]{acmart}
\AtBeginDocument{%
  }

\usepackage{xspace}
\usepackage{pifont}
\usepackage{booktabs}
\usepackage[most]{tcolorbox}
\usepackage{listings}
\usepackage{enumitem}
\usepackage{multirow}
\usepackage{makecell}
\newcommand*\name{\textsc{c-CRAB}\xspace}

\newcommand{\rqans}[2]{
    \begin{tcolorbox}[
        left=5pt, right=2pt, top=1pt, bottom=1pt,
        boxrule=0pt,
        frame hidden,
        sharp corners,
        enhanced,
        borderline west={2.5pt}{0pt}{gray},
        colback=white,
    ]
    \textbf{#1}~{#2}
    \end{tcolorbox}
}
\newcommand{\quotebox}[1]{
    \begin{tcolorbox}[
        left=5pt, right=2pt, top=1pt, bottom=1pt,
        boxrule=0pt,
        frame hidden,
        sharp corners,
        enhanced,
        borderline west={2.5pt}{0pt}{gray},
        colback=white,
    ]
    {#1}
    \end{tcolorbox}
}

\newcommand\arxivurl{\url{https://github.com/c-CRAB-Benchmark}}

\lstdefinestyle{mycode}{
  language=Python,
  basicstyle=\ttfamily\footnotesize,
  breaklines=true,
  breakatwhitespace=false,
  columns=fullflexible,
  keepspaces=true,
  showstringspaces=false,
  frame=single,
  rulecolor=\color{black!25},
  backgroundcolor=\color{black!2},
}

\renewcommand\footnotetextcopyrightpermission[1]{}

\usepackage{fancyhdr}

\fancypagestyle{plain}{%
  \fancyhf{}                  
  \fancyfoot[C]{\thepage}     

}

\begin{document}

\title{Code Review Agent Benchmark}

\author{Yuntong Zhang}
\authornote{The first two authors contributed equally to this research.}
\email{zhang.yuntong@u.nus.edu}
\orcid{0009-0005-1664-7110}
\affiliation{%
\institution{National University of Singapore}
\country{~}
 }

\author{Zhiyuan Pan}
\orcid{0009-0006-6059-5191}
\authornotemark[1]
\authornote{Work done while author is full-time at National University of Singapore.}
\email{zy\_pan@zju.edu.cn}
\affiliation{
 \institution{Zhejiang University}
\country{~}
}

\author{Imam Nur Bani Yusuf}
 \email{inbyusuf@nus.edu.sg}
 \authornote{Corresponding author.}
\orcid{0000-0003-0900-5230}
 \affiliation{%
   \institution{National University of Singapore}
 \country{~}
 }

 \author{Haifeng Ruan}
 \email{haifeng.ruan@u.nus.edu}
 \orcid{0009-0008-1080-4770}
 \affiliation{%
   \institution{National University of Singapore}
 \country{~}
 }

\author{Ridwan Shariffdeen}
 \email{shariffdeenr@acm.org}
 \orcid{0000-0001-5409-4864}
 \affiliation{%
   \institution{SonarSource}
\country{~}
 }

\author{Abhik Roychoudhury}
 \email{abhik@nus.edu.sg}
\orcid{0000-0002-7127-1137}
 \affiliation{%
   \institution{National University of Singapore}
\country{~}
 }

\renewcommand{\shortauthors}{Zhang et al.}

\begin{abstract}
Software engineering agents have shown significant promise in coding tasks - either writing an entire application, or writing a code patch to improve a software project by remediating an issue. As AI agents permeate code writing, and generate huge volumes of code automatically - the matter of code quality comes front and centre. As the automatically generated code gets integrated into huge code-bases — the issue of code review and broadly quality assurance becomes important. In this paper, we take a fresh look at the problem and curate a code review dataset for AI agents to work with. Our dataset called \name (pronounced see-crab) can evaluate agents for code review tasks. Specifically given a pull-request (which could be coming from code generation agents or humans), if a code review agent produces a review, our evaluation framework can asses the reviewing capability of the code review agents. Our evaluation framework is used to evaluate the  state of the art today - the open-source PR-agent, as well as commercial code review agents from Devin, Claude Code, and Codex.  

Our \name dataset is systematically constructed from human reviews - given a human review of a pull request instance 
we generate corresponding tests to evaluate the code review agent generated reviews. Such a benchmark construction gives us several insights. First of all, all of the existing review agents taken together can solve only around ~40\% of the \name tasks, indicating the potential to close this gap by future research. Secondly, we observe that the agent reviews often consider different aspects from the human reviews - indicating the potential for human-agent collaboration for code inspection and review that could be deployed in future software teams. Last but not the least, the agent generated tests from our data-set act as a held out test-suite and hence quality gate for agent generated reviews. What this will mean for future collaboration of code generation agents, test generation agents and code review agents - remains to be investigated. 
\end{abstract}



\maketitle

\pagestyle{plain}

\section{Introduction}

Code review is a fundamental quality assurance practice in modern software development. Before a proposed change is merged into the main codebase, developers inspect pull requests (PRs) to detect defects, enforce project standards, and maintain long-term code quality. Traditionally, this process relies on human reviewers who carefully examine pull requests (PRs) and provide feedback to authors. 
Recent progress in software engineering agents has significantly accelerated the process of writing code. AI systems can now generate large amounts of code automatically, both for implementing new features~\cite{DBLP:conf/acl/LiZGM0PHWL25, zhou2026featurebenchbenchmarkingagenticcoding} and for resolving issues~\cite{DBLP:conf/acl/XieLGDLZ025, DBLP:conf/issta/0002RFR24} in existing repositories. As these systems become increasingly integrated into development workflows, they contribute to a growing number of generated code and submitted PRs~\cite{DBLP:journals/corr/abs-2601-18341, DBLP:journals/corr/abs-2601-17627}. However, the capacity of human reviewers has not scaled at the same pace~\cite{DBLP:journals/corr/abs-2509-06216}, creating a new bottleneck in the development pipeline: while code generation is increasingly automated, ensuring the quality of generated code still relies heavily on manual inspection.

This challenge has motivated the development of automated code review agents that analyze pull requests and generate review feedback~\cite{DBLP:conf/emnlp/TangKSLLETKB24, DBLP:journals/corr/abs-2404-18496, DBLP:journals/corr/abs-2511-00517, DBLP:journals/corr/abs-2601-01129}. Several tools and systems have recently been proposed to automatically produce review comments aimed at identifying bugs, design issues, or maintainability concerns. Yet a central question remains: \emph{to what extent do these automated review tools identify the same issues that human reviewers would raise?}

Existing evaluation methods do not answer this question well. Most benchmarks compare generated comments against human-written reviews using textual overlap or embedding similarity metrics~\cite{DBLP:conf/icse/TufanoPTPB21, DBLP:conf/icse/TufanoMMPPB22, DBLP:conf/naacl/NaikAFR25, DBLP:journals/corr/abs-2501-05176, DBLP:journals/corr/abs-2511-07017}. Such metrics primarily measure resemblance in wording rather than whether a review identifies a meaningful issue in the code. However, we posit that these evaluation approaches are not well suited for measuring how closely AI-generated reviews align with the concerns raised by human reviewers. Human review comments are often noisy artifacts of the review process: they may include clarification questions, subjective stylistic suggestions, or back-and-forth discussions between reviewers and authors. 
As a result, the wording of a comment does not always correspond directly to the underlying issue being raised. 
Measuring similarity between comments is therefore problematic. 
A review may correctly identify a valid issue using entirely different wording and still receive a low similarity score. 
Similar limitations also arise in other evaluation methods, such as localization metrics~\cite{DBLP:journals/corr/abs-2509-14856, zhang2026aacrbenchevaluatingautomaticcode} and LLM-as-a-judge~\cite{DBLP:journals/corr/abs-2509-14856}. Localization metrics only assess whether an issue is identified at the same code location as the ground truth human reviews while ignoring the substance of the review. Furthermore, LLM-as-a-judge approaches can be sensitive to prompt design and inherent randomness, causing their judgments difficult to reproduce reliably.


In this paper, we revisit the problem of evaluating automated code review tools and introduce \textbf{\name} (pronounced \emph{see-crab}), a benchmark designed to assess how well automated review tools identify the same issues that human reviewers raise in realistic PR settings. 
Instead of evaluating reviews solely based on similarity to reference comments, \name adopts a test-based evaluation. 
Specifically, human review feedback is systematically converted into executable tests that capture the underlying issues identified during the review process. These tests serve as objective evaluation oracles: if a review agent identifies an issue corresponding to a failing test, the issue can be verified automatically in the code. 

To evaluate a review tool, we provide the tool with a PR and collect the review comments it produces. A separate coding agent then revises the code in the PR based on the generated review comments from the review tools, and the revised version is executed against the curated tests. If the tests pass, this indicates that the review successfully identifies actionable issues that lead to correct code improvement. By doing so, \name measures the extent to which automated review tools discover issues originally identified by human reviewers. This evaluation therefore measures whether automated reviewers can raise issues that correspond to those raised by human reviewers in real PRs.

Using \name, we evaluate several state-of-the-art review agents, including the open-source PR-Agent, and the review capabilities of Devin Review, Claude Code, and Codex. 
Our results show that current systems successfully identify only about 40\% of the issues captured in the benchmark, indicating substantial room for improvement in automated code review. 
We further observe that AI-generated reviews often focus on different aspects of code compared to human reviewers.
However, they remain valuable in identifying problems and suggesting improvements.
This difference highlights complementary perspectives that may support future human-AI collaborative review workflows. 
In summary, the contributions of this paper are as follows:

\begin{itemize}[leftmargin=*]
\item We introduce \textbf{\name}, a benchmark dataset for evaluating automated code review agents on real-world pull requests.
\item We propose an evaluation framework that converts human review feedback into executable tests, enabling objective validation of review quality.
\item We evaluate several state-of-the-art review agents and show that current systems solve only about 40\% of the benchmark tasks, indicating substantial room for improvement in automated code review.
\item We analyze differences between human and AI-generated reviews, revealing complementary perspectives that motivate future human-agent collaboration in code review.
\end{itemize}

\section{Background and Related Work}
\label{sec:related-work}

\begin{table*}[t]
\centering
\caption{Comparison of automated code review benchmarks.}
\label{tab:related-work}
\small
\begin{tabular}{p{2.5cm} c p{3.5cm} c p{5.5cm}}
\toprule
\multicolumn{1}{c}{\textbf{Dataset}} &
\multicolumn{1}{c}{\textbf{Year}} &
\multicolumn{1}{c}{\textbf{Oracle source}} &
\multicolumn{1}{c}{\textbf{Repo-level Support}} &
\multicolumn{1}{c}{\textbf{Evaluation}} \\
\midrule
Tufano-21~\cite{DBLP:conf/icse/TufanoPTPB21}
& 2021
& Mined OSS PR reviews
& No
& N-gram overlap \\

CodeReviewer~\cite{DBLP:conf/sigsoft/LiLGDJJMGSFS22}
& 2022
& Mined OSS PR reviews
& No
& N-gram overlap \\

Tufano-22~\cite{DBLP:conf/icse/TufanoMMPPB22}
& 2022
& Mined OSS PR reviews
& No
& N-gram overlap \\

CRScore corpus~\cite{DBLP:conf/naacl/NaikAFR25}
& 2025
& LLM + static analysis
& No
& Embedding similarity \\

ContextCRBench~\cite{DBLP:journals/corr/abs-2511-07017}
& 2025 
& Mined OSS PR reviews 
& No
& N-gram overlap \\

SWE-CARE~\cite{DBLP:journals/corr/abs-2509-14856}
& 2025
& Mined OSS PR reviews
& Yes
& Localization + N-gram overlap + LLM-as-a-judge \\

AACR-Bench~\cite{zhang2026aacrbenchevaluatingautomaticcode}
& 2026
& Mined OSS PR reviews
& Yes
& Localization \\

RovoDev~\cite{DBLP:journals/corr/abs-2601-01129}
& 2026
& Logs + user feedback
& Yes
& Online production metrics \\

Qodo Benchmark~\cite{yanay2026qodoBenchmarkBlog}
& 2026
& Mined OSS PR reviews
& Yes
& LLM-as-a-judge \\

CR-Bench~\cite{pereira2026crbenchevaluatingrealworldutility}
& 2026
& Mined OSS PR issue description
& Yes
& LLM-as-a-judge \\

\name (Ours)
& 2026
& Mined OSS PR reviews
& Yes
& Test-based \\
\bottomrule
\end{tabular}
\end{table*}

\subsection{Automated Code Review Tools}

Automated code review tools can be broadly categorized into static analysis-based approaches and AI-based approaches. 
Static analysis tools detect vulnerabilities and code quality issues based on pre-defined rules, patterns, or data-flow analysis. 
In contrast, AI-based review tools typically leverage LLMs to detect issues based on learned patterns during LLM's training, as well as contextual knowledge provided through prompts or agentic scaffold.
These review tools are typically triggered when a Pull Request (PR) is opened on GitHub, GitLab, or BitBucket. 
When triggered, AI-based review tools collect necessary context, such as the modified lines in the PR, the surrounding code context, and repository-level context such as dependency graphs.
This collected context is then analyzed by an LLM to identify potential issues. 
Some AI-based review tools may perform additional post-processing, such as false positive filtering or severity ranking, before the reviews are presented to developers.
As output, most review tools present the reviews as a summary comment or inline comments in the PR.
In this paper, we focus specifically on the evaluation of AI-based code review tools.

\subsection{Benchmarks for Automated Code Review}
Table~\ref{tab:related-work} summarizes the existing available benchmarks for code review. 
Early work primarily focuses on fine-grained settings, where review comments are aligned with localized code changes such as lines, methods, or diff hunks. Datasets, such as in~\cite{DBLP:conf/sigsoft/LiLGDJJMGSFS22, DBLP:conf/icse/TufanoPTPB21, DBLP:conf/icse/TufanoMMPPB22, DBLP:conf/naacl/NaikAFR25, DBLP:journals/corr/abs-2511-07017}, construct large-scale pairs of code changes and human-written comments mined from open-source repositories. These datasets have been widely used to train and evaluate models for review comment generation.
However, such datasets typically have limited context granularity, often isolating individual diff hunks and abstracting away the broader pull requests and repository context. Consequently, they do not fully capture the complexity of real-world code review.
More recent benchmarks move toward PR-level and repository-aware evaluation. Datasets, such as in~\cite{DBLP:journals/corr/abs-2509-14856, zhang2026aacrbenchevaluatingautomaticcode, DBLP:journals/corr/abs-2601-01129}, incorporate richer contextual signals, including full pull requests, surrounding code, and repository structure. These benchmarks better reflect practical deployment settings, where review tools must reason over larger contexts and diverse types of issues.

Despite these progress, existing benchmarks differ in how they define evaluation oracles. Most rely on mined human review comments as reference signal, while more recent work introduces expert annotations~\cite{zhang2026aacrbenchevaluatingautomaticcode} or LLM-augmented labels~\cite{DBLP:conf/naacl/NaikAFR25} to improve data quality. Nevertheless, the majority of benchmarks still treat natural-language human comments as the primary ground truth, which introduces challenges due to noise, incompleteness, and variability in human review practices. Ours is the only dataset that depends on dynamic program behavior as witnessed by tests (albeit generated from human reviews).




\subsection{Evaluation Methods for Code Reviews}
A common strategy for evaluating automated code review systems compares generated comments against human-written reviews based on similarity with  the human reviews, such as BLEU~\cite{DBLP:conf/acl/PapineniRWZ02}, ROUGE~\cite{lin-2004-rouge}, chrF~\cite{DBLP:conf/wmt/Popovic15}, and embedding-based similarity~\cite{DBLP:conf/icse/TufanoPTPB21, DBLP:conf/icse/TufanoMMPPB22, DBLP:journals/corr/abs-2501-05176, DBLP:conf/naacl/NaikAFR25, DBLP:journals/corr/abs-2511-07017}.
While these metrics are easy to compute at scale, 
recent studies further show that human review comments used as references can be noisy or incomplete~\cite{DBLP:conf/fase/LuLHYCYZZ25,DBLP:journals/corr/abs-2501-05176}.

Several studies also uses localization to evaluate code review tools~\cite{DBLP:journals/corr/abs-2509-14856, zhang2026aacrbenchevaluatingautomaticcode}. This metric measures whether automated code review tools identify concerns at the same locations as human reviewers. However, it evaluates only the location of identified issues, instead of the content or nature of the concerns raised. Another study~\cite{DBLP:journals/corr/abs-2601-01129} evaluates code review tools using online production metrics, such as acceptance rate. While these metrics more accurately reflect real-world usage, they are difficult to obtain because they require access to a production environment for data collection.

Another emerging paradigm uses large language models as judges to directly assess review quality~\cite{DBLP:journals/corr/abs-2509-01494,DBLP:journals/corr/abs-2509-14856}. In this setting, an LLM is prompted to assess the quality of generated review comments by comparing them with human-written feedback or by reasoning about the code changes and the identified issues~\cite{DBLP:journals/corr/abs-2509-01494,DBLP:journals/corr/abs-2509-14856}. Compared with lexical or embedding-based similarity metrics, LLM-based evaluation can consider richer contextual information and may better capture aspects such as correctness, relevance, and usefulness of review feedback. Although promising, LLM-as-judge approaches can suffer from bias, instability, and sensitivity to prompt design~\cite{DBLP:conf/ijcnlp/ShiMLDMV25,DBLP:journals/corr/abs-2406-12624}. These factors make it difficult to ensure reproducibility and consistency when LLMs are used as evaluation oracles.

In contrast, \name introduces a test-based evaluation framework for automated code review. Instead of comparing generated reviews against reference comments, we convert human review feedback into executable tests that capture the underlying issues identified by humans during the review process. Each benchmark instance includes an executable environment, enabling automated reviewers to be evaluated in a reproducible setting. These tests serve as objective evaluation oracles, allowing scalable, stable, and verifiable assessment of automated review agents.

\section{Motivation}
\label{sec:motivating-example}
First, we demonstrate that existing evaluation methods fail to capture whether automated reviews align with the intentions of human reviewers. We then show how our test-based evaluation in \name overcomes this limitation.

\subsection{Comparison with Similarity Metrics}
Consider the pull request from \texttt{python-telegram-bot}\footnote{\url{https://github.com/python-telegram-bot/python-telegram-bot/pull/3514\#discussion_r1083264889}}, which introduces a check to reject keyboard inputs with more than two dimensions.

\begin{figure}[ht]
\begin{lstlisting}[style=mycode]
if isinstance(keyboard[0][0], Sequence) and not isinstance(keyboard[0][0], (str, bytes))
\end{lstlisting}
\caption{The line of code under review.}
\label{fig:reviewed-line}
\end{figure}

A human reviewer raised the following concern on the line shown in Figure~\ref{fig:reviewed-line}:
\quotebox{
\small
At this point we haven't yet established that the rows are sequences as well. \texttt{check\_keyboard\_type([1])} would raise an exception with this addition IISC.
}
\noindent This comment identifies a robustness issue: the implementation assumes that nested elements are sequences and may raise an exception on malformed inputs (e.g., \texttt{[1]}) instead of safely rejecting them.

Now consider the review generated by Codex for the same patch:
\quotebox{
\small
\texttt{keyboard[0][0]} is accessed before validating shape/contents, so malformed but common invalid inputs like \texttt{[]}, \texttt{[[]]}, or \texttt{[1]} can raise \texttt{IndexError}/\texttt{TypeError} instead of returning \texttt{False} (and then producing the expected \texttt{ValueError} in markup constructors).
}
\noindent We can observe that this comment raises the same concern about indexing that can lead to runtime errors. However, the two comments describe the issue in different ways.

\begin{table}[h]
\centering
\small
\caption{Comparison of \name against prior similarity-based metrics.}
\label{tab:motivating-similarity}
\begin{tabular}{l r p{4cm}}
\toprule
\textbf{Metric} & \textbf{Value} & \textbf{Interpretation} \\
\midrule

 BLEU-4   & 0.00  & \multirow{3}{4cm}{The degree of n-gram overlap between human and automated tool reviews.} \\
 ROUGE-L  & 7.02  & \\
 chrF     & 20.74 & \\[1ex]

\midrule

 Embedding similarity & 54.59 & The relative distance between human and automated tool reviews in the embedding space. \\

\midrule

\multicolumn{3}{l}{\textit{Our Work}} \\

\quad Test passed? & \textit{Yes} & Whether the reviews successfully capture concerns raised by humans during review. \\

\bottomrule
\end{tabular}
\end{table}

Table~\ref{tab:motivating-similarity} reports the similarity scores obtained using metrics commonly adopted in prior work on code review evaluation. Here, BLEU-4 measures the precision of up to 4-gram overlaps with a brevity penalty, ROUGE-L captures the longest common subsequence (LCS) to reflect sequence-level similarity, and chrF computes an F-score based on character-level 6-gram overlap. For the embedding similarity, we convert both reviews into an embedding vector using SentenceTransformers\footnote{\url{https://sbert.net}} and compute the cosine similarity, following the prior work~\cite{DBLP:journals/corr/abs-2501-05176}. All scores are normalized to 0--100. 

The result show that all similarity metrics yield low scores. 
The low scores on BLEU-4, ROUGE-L, and chrF are expected due to the minimal n-gram overlap between the human and automated tool reviews. The embedding-based similarity yields better score than chrF. However, the score remains moderate and cannot precisely capture that both reviews raise the same concern. This is because embedding-based similarity measures the proximity in the embedding space and the embedding representation is affected by how the reviews are expressed (e.g., word choices, sentence structure)~\cite{DBLP:journals/tacl/AntoniakM18}. As a result, two reviews that describes the same concern can still be relatively distant in the embedding space if they differ in phrasing.


\subsection{Comparison with LLM-as-a-Judge}

\begin{figure}[ht]
\begin{lstlisting}[style=mycode]
live_events_columns: ArrayField = ArrayField(models.CharField(max_length=200), null=True, blank=True)
\end{lstlisting}
\caption{The line of code under review.}
\label{fig:reviewed-line2}
\end{figure}

Let us consider another pull request from \texttt{posthog}\footnote{\url{https://github.com/posthog/posthog/pull/10293\#discussion\_r897800551}}. As shown in Figure~\ref{fig:reviewed-line2}, a field with type \texttt{ArrayField} was added to a class. 
On this line of code change, a human reviewer raised the concern:
\quotebox{
\small
Let's make it a \texttt{TextField}. Somebody might want to put three 60-character-long columns here, and boom, we're out of space.
}
\noindent Claude Code generated the following review comment:
\quotebox{
\small
posthog/models/team/team.py:line 114
\texttt{live\_events\_columns} uses \texttt{ArrayField(models.CharField(max\_length=200))}. Property keys can be arbitrary strings; a hard limit of 200 characters may silently truncate very long property names or cause save failures. Consider aligning this with \texttt{person\_display\_name\_properties} which uses \texttt{max\_length=400}, or document the constraint.
}

Claude Code does a good job at defect detection and points out the same issue as human review: \texttt{CharField} in \texttt{django} library cannot handle arbitrarily long strings. However, the automated review comment is still not fully aligned with the human reviewer's intent. Here, its suggested fix is different: it only proposes a trivial mitigation while the human reviewer suggested replacing \texttt{CharField} with \texttt{TextField}, which truly solves the root issue. 

For prior work that adopts vanilla LLM-as-a-Judge evaluation, LLM is typically used to judge whether an automated review points out the same issue as human reviews. However, as the above example shows, alignment with human intent means more than this, as the correctness and quality of suggested fix also plays an important role in code review, and previous LLM-as-a-Judge methods cannot distinguish between the above two review comments.

\subsection{Our benchmark: \name} 
The key idea of \name is to evaluate review comments based on whether they lead to behaviorally correct fixes. During evaluation, we provide the generated review to a coding agent, which revises the patch accordingly. The revised patch is then executed against the synthesized test. If the revised patch passes the test, we consider the review useful, as it successfully captures the underlying issue originally identified by the human reviewer.

For the human review in Figure~\ref{fig:reviewed-line}, we convert the human review into the executable test shown in Figure~\ref{lst:keyboard-row-type-regression} to capture its intent.
\begin{figure}[t]
\begin{lstlisting}[style=mycode]
def test_check_keyboard():
    """
    Test for a bug where check_keyboard_type tries to access keyboard[0][0] before validating that rows are sequences. For keyboard=[1], the function should not raise an exception and should return False.
    """
    keyboard = [1]
    try:
        result = check_keyboard_type(keyboard)
    except Exception as exc:
        ...
    assert result is False
\end{lstlisting}
\caption{The test ensuring \texttt{check\_keyboard\_type([1])} returns \texttt{False} instead of raising an exception.}
\label{lst:keyboard-row-type-regression}
\end{figure}
This test directly captures the reviewer's concern by asserting that the function must not perform unsafe nested indexing before validating input structure. Unlike similarity-based metrics, which rely on textual overlap or semantic similarity of text, this test encodes the expected behavior as an executable specification. In this example, the review generated by the automated review tool enables the coding agent to revise the initial patch that passes the test. This indicates that, despite low textual similarity, the generated review successfully captures the same underlying issue as the human review and is therefore considered useful under our evaluation.

For the human review in Figure~\ref{fig:reviewed-line2}, the tool-generated review does not successfully guide the coding agent toward a patch that satisfies the executable test. This is because the test expects the presence of \texttt{TextField} in the source code, whereas the tool-generated review proposes an alternative fix by increasing the maximum length. Consequently, the revised patch fails to satisfy the test. In contrast, in our experiment, an LLM-as-a-judge considers both reviews to raise the same concern although both reviews suggesting different fixes. This ability to capture the \emph{precise} review intent using executable tests is what makes \name different compared to LLM-as-a-judge. Moreover, another concern with the LLM-as-a-judge is the inherent randomness of its output that makes the evaluation not reliable.
\section{Methodology}
\label{sec:methodology}
In this section, we first present the high-level design of \name, followed by discussion on benchmark construction, evaluation procedure, and implementation details.


\subsection{\name Overview}
The challenge in evaluating code review lies in the fact that the value of a review comment is not determined by how closely it matches human wording, but by whether it identifies an issue whose resolution improves the code. This observation motivates the design of \name, a benchmark that evaluates review comments based on whether they lead to verifiable improvements in the code.

At a high level, each benchmark instance consists of a pull request containing a patch to be reviewed and a set of test cases derived from human review feedback. 
These tests encode issues identified by human reviewers. 
Each test is constructed to fail on the original patch and pass once the corresponding issue has been resolved.
During evaluation, an automated review tool is provided with a PR instance and asked to analyze the PR and generate review comments. 
These comments are used as guidance for a coding agent to revise the code changes in the PR. 
The revised code is subsequently executed against the benchmark tests. 
Because each test represents an issue originally identified by human reviewers, passing a test indicates that the corresponding issue has been successfully addressed. 
Therefore, if a test fails on the original PR but passes after the revision, we attribute the improvement to the usefulness of the review feedback that guided the modification. 
\name\ measures review usefulness by whether generated feedback leads to resolving human-identified issues, as shown by passing of corresponding tests.

\subsection{Benchmark Curation}
\label{sec:benchmark-curation}

\begin{figure}[t]
    \centering
    \includegraphics[width=\linewidth]{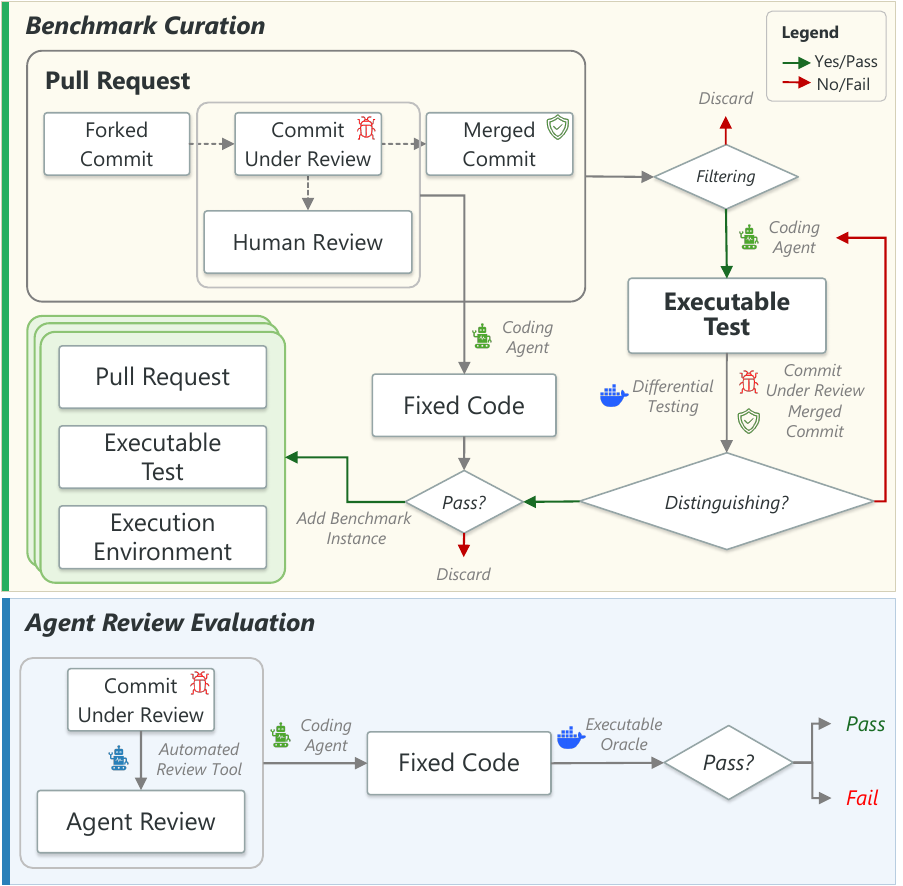}
    \caption{Idea behind Benchmark Construction.}
    \label{fig:idea}
\end{figure}

Figure~\ref{fig:idea} illustrates our benchmark curation pipeline. For each pull request, the pipeline takes three artifacts as input: the pull request description, the candidate patch, and the associated human review comments. The goal of the pipeline is to construct a benchmark instance that preserves the PR context while providing a set of tests derived from human reviews. 
This goal is nontrivial for two reasons. First, raw code review discussions are inherently noisy: many comments are conversational or exploratory and therefore cannot be reliably translated into executable tests. 
Second, generated tests must be executed within a properly constructed environment that reproduces the repository’s dependencies and runtime conditions. Without such infrastructure, tests cannot be executed consistently during evaluation.

To address these challenges, we design a structured curation pipeline consisting of four stages: \emph{review filtering}, \emph{executable environment construction}, \emph{converting NL comments to executable tests}, and \emph{validation with coding agents}. First, we filter review comments to retain only those that correspond to objectively verifiable issues. Second, we construct an executable environment for each PR instance to ensure that generated tests can be executed reliably. Third, we synthesize tests from the retained review comments. 
The key idea is to encode the issue raised by a human review comment as a test that fails on the candidate patch but passes once the issue is fixed. 
Finally, we perform additional validation to ensure that benchmark instances contain only comments that can be resolved by a coding agent.
We describe each stage of the pipeline below.

\subsubsection{Review Filtering}
\label{sec:filtering}
The objective of the filtering stage is to retain only review comments that can be translated into high-quality test cases serving as benchmark targets. In raw PR discussions, not every human comment reflects a concrete and actionable concern in the candidate patch. Many comments are conversational in nature, such as requests for clarification, acknowledgments, or praise. While such comments may still play an important role in real-world collaboration, they are not suitable for our benchmark because they cannot be mapped unambiguously to executable tests. 
Furthermore, these types of comments are generally not expected to be produced by automated code review tools. Including them would introduce noise into the curation pipeline and reduce the validity of the resulting tests. Therefore, the goal of the filtering stage is to identify review comments that express verifiable issues whose resolution can, in principle, be validated through execution. 

To achieve this in a scalable manner, we design an automated filtering classifier by prompting an LLM. 
To define the filtering prompt, we first construct a small gold set $\mathcal{D}$ through manual annotation, which is used to develop and refine the LLM prompt later. 
We randomly sample 100 review comments from the candidate pool, and two authors independently label each comment as either {high-quality} or {low-quality} for benchmark construction, providing a short justification for each label. A comment is labeled as high-quality if it identifies a specific, actionable, and objectively verifiable issue in the patch. Conversely, comments are labeled as low-quality if they lack a concrete action item, such as conversational exchanges, acknowledgments, or concerns that cannot be reliably validated through testing. After the independent annotation phase, the two authors compare their labels and resolve any disagreements through discussion, resulting in a set of ground truth labels for $\mathcal{D}$.

Given this gold set with ground-truth labels, we iteratively refine an LLM prompt so that an LLM can make correct classifications in the gold set. 
This prompt refinement process continues until the prompt can make the LLM-based classification reach a predefined precision threshold. 
The resulting prompt is used to classify the remaining review comments outside of the gold set, filtering out low-quality instances.

\subsubsection{Execution Environment Construction}

An execution environment is essential in the curation of \name, as a generated test has to be executed and verified before being used as a reliable oracle for evaluating review comments. During this stage, we construct an isolated Docker environment for each pull request in the dataset.
A successfully built environment is a Docker image that consists of pre-installed operating system and software dependencies, and a pre-configured source code repository, providing out-of-the-box scaffolding for subsequent pipeline stages.

The detailed steps to build such environments are as follows. First, we generate scripts that clone the git repository and check out the commit to be reviewed. Second, we implement a heuristic script builder that automatically generates installation scripts by detecting build tools and inferring software dependencies and their versions from the corresponding specification files (e.g., \texttt{setup.py} and \texttt{pyproject.toml} in Python projects). These scripts, along with Dockerfiles that set up the operating system, form all necessary components required by execution environment construction. Finally, each built Docker image is tested to ensure that an editable installation of the repository exists and all dependencies are installed.

In some cases, especially when it comes to historic pull requests, the automatically generated installation scripts can fail because the original specifications are too loose now. 
For example, some projects did not specify the versions or only specified lower bounds for some of the dependencies. 
To overcome the issue for these historic commits, we use an automated coding agent running in an isolated environment to audit the dependency requirements, pinpoint the accurate dependency versions and override the previously generated installation script. 

\subsubsection{Converting NL Comments to Tests}
\label{sec:test-gen}
The objective of this stage is to transform each retained natural-language review comment into an \emph{executable test} that captures the issue raised by the human reviewer. 
For a review comment $c$ associated with file $f$, the input to this stage consists of three components: (1) the natural-language comment and its corresponding diff hunk, (2) the repository state at the PR head commit (the \texttt{before} version), and (3) the repository state after the issue has been addressed (the \texttt{after} version). Given this information, we invoke an LLM-based loop to synthesize a test case that reflects the issue described in the comment. The goal is to generate a test case $t_c$ that fails in the \texttt{before} version and passes in the \texttt{after} version.
The generated test $t_c$ therefore serves as an executable oracle representing the issue originally identified by the human reviewer.

Before generating tests, we construct the contextual information required to interpret each review comment. Review comments in pull requests often refer to specific lines within a diff.
Therefore, we align each comment with its corresponding code region using the diff metadata provided by the pull request. Then, we retrieve the full file contents of the affected file $f$ at both the before and after commits
This context provides the LLM with sufficient information to understand the intended behavior and the modification that resolves the issue.

Using the extracted context, we prompt an LLM to generate a candidate test case that captures the issue described in the comment. The prompt includes the review comment, the relevant diff hunk, and the file-level code context. After a candidate test is generated, we execute it against both the before and after repository states. The execution outcome determines whether the test satisfies the fail-then-pass requirement. If the test fails on the before version and passes on the after version, it is accepted as a valid oracle.

In practice, the initial generation may not satisfy the fail-then-pass requirement due to issues such as incorrect imports, runtime errors, or assertions that do not distinguish between the two repository states. To address this, we employ an execution-guided refinement loop. When a generated test does not satisfy the required behavior, the execution traces and error messages are fed back to the coding agent, which revises the test accordingly. This process continues for a fixed number of iterations until either a valid test is produced or the attempt budget is exhausted.

\subsubsection{Validation with Coding Agents}
\label{sec:curation-validation}
The final stage of the curation pipeline ensures that each benchmark instance can be reliably resolved by the coding agent used in the evaluation phase when provided with the corresponding human review. This validation step is necessary because the evaluation protocol of \name\ relies on a coding agent to modify the patch according to generated review comments. Therefore, we must ensure that when the coding agent fails to make the generated test pass during evaluation, the failure can be attributed to the quality of the review feedback rather than limitations of the coding agent itself.

Concretely, for each instance, we ask a coding agent $\mathcal{A}$ to improve the code using the human review comment as guidance. The agent is provided with the patch under review, the relevant file context, and the natural-language review comment that motivated the generated test. The tests produced during the test generation stage are not exposed to the agent and are used only for post-hoc verification. An instance is considered valid if the agent produces a revision that causes the corresponding test to pass, thereby satisfying the fail-then-pass property. If the agent fails to produce a revision that passes the test within a fixed number of attempts, the instance is discarded.

\subsection{Benchmark Usage}
\subsubsection{Workflow}
\label{sec:eval-workflow}
The goal of the evaluation phase is to measure the usefulness of review feedback generated by automated code review tools. Specifically, we evaluate whether the issues raised by an automated review tool can guide the coding agent $\mathcal{A}$ (the same coding agent used in Section~\ref{sec:curation-validation}) to improve the patch under review and cause the executable tests to pass.

The evaluation comprises two stages: review generation and review-guided revision. In the first stage, each automated review tool analyzes every instance in the final benchmark and generates review comments given the patch under review and PR description.
In the second stage, the generated reviews are used to guide the coding agent $\mathcal{A}$ to revise the patch. Specifically, we provide the agent with the patch under review and the review comments produced by the tool under evaluation. The agent attempts to modify the patch according to the generated reviews. The revised patch is then executed against the tests associated with the benchmark instance. If the revision causes a test to pass, the corresponding review feedback is considered aligned with human reviews, as it successfully guides the same agent $\mathcal{A}$ to resolve the issue originally identified by human reviewers.

\subsubsection{Evaluation Metric} 
Formally, let $T_i$ denote the validated test set associated with benchmark instance $i$, and let $\hat{P}_i$ denote the patch produced by the coding agent $\mathcal{A}$ when guided by the review findings generated by a tool. We define the instance-level performance as the pass rate of the validated tests:
\[
s_i=\frac{|\{t \in T_i \mid t(\hat{P}_i)=\texttt{PASS}\}|}{|T_i|}.
\]
The overall performance of a review tool is measured by the aggregate test pass rate across benchmark instances. 
We note that the pass rate in \name reflects the extent to which review tools can identify issues found by human reviewers (i.e., the true positive rate when human reviews are treated as ground truth).
Automated review tools may also generate other valuable comments that are not identified by human reviewers; however, like other existing benchmarks, \name does not directly evaluate these additional comments.
To provide a more complete picture, we further discuss the quality of such additional reviews generated by automated tools in Section~\ref{sec:rq1}.


\subsection{Benchmark Implementation}

We build \name\ on top of SWE-CARE~\cite{DBLP:journals/corr/abs-2509-14856} because it already provides PR instances with the necessary commit metadata
More generally, however, our pipeline is dataset-agnostic: any pull request that provides the required metadata could serve as a candidate instance for the curation process. 

\subsubsection{Implementation of the Curation Pipeline} 
We leverage GPT-5.2 to perform review filtering (Section~\ref{sec:filtering}).
GPT-5.2 is also used for NL comments to tests conversion (Section~\ref{sec:test-gen}), and up to three LLM attempts with execution feedback are allowed to generate the final test.
For validation (Section~\ref{sec:curation-validation}), we use Claude Code with Sonnet-4.6 backend as the coding agent. 


\subsubsection{Benchmark Statistics}

\begin{table}[t]
\centering
\small
\caption{The number of remaining PR instances and Review Comments after each processing step.}
\begin{tabular}{l r r r r}
\toprule
Pipeline Stage & \# PR & \# Comments  \\
\midrule
Initial Dataset & 671 &  1,313  \\
Review Filtering & 410 &  595  \\
Executable Environment Construction & 410 &  595  \\
Converting NL Comments to Tests & 339 & 481  \\
Validation with Coding Agents \textbf{(Final)} & 184 & 234  \\
\bottomrule
\end{tabular}
\label{tab:pipeline}
\end{table}


\begin{table}[t]
\centering
\small
\caption{Statistics of the final \name\ benchmark.}
\label{tab:stats}
\begin{tabular}{lr}
\toprule
\textbf{Statistic} & \textbf{Value} \\
\midrule
\multicolumn{2}{l}{\textit{Pull Requests}} \\
\quad Repositories & 67 \\
\quad Instances (PRs) & 184 \\
\quad Avg.\ modified lines & 418.1 \\
\midrule
\multicolumn{2}{l}{\textit{Tests / Review Comments}} \\
\quad Total review comments (tests) & 234 \\
\qquad Behavioral & 42 (17.9\%) \\
\qquad Structural & 192 (82.1\%) \\
\quad Avg.\ tests per instance & 1.27 \\
\quad Avg.\ lines per test & 31.8 \\
\bottomrule
\end{tabular}
\end{table}

Table~\ref{tab:pipeline} summarizes the number of remaining pull requests and review comments after each stage of the curation pipeline. 
Overall, the resulting benchmark contains 184 pull request instances and 234 validated review comments, each associated with at least one executable test. 
The reduction in number of comments from initial dataset is expected, given the strict requirements imposed at each stage of the pipeline.
The final dataset represents a carefully curated subset of the original data and supports a faithful evaluation of review usefulness.
The detailed statistics of the curated benchmark is shown in Table~\ref{tab:stats}.
The 234 tests in \name are further categorized into \textit{behavioral} and \textit{structural}.
Behavioral tests import and execute the tested code at runtime.
They invoke the tested functions with specific inputs, and check outputs or verify exceptions.
On the other hand, structural tests inspect source code text, matches patterns, and check API surfaces to determine whether desired code changes have been made.

\subsubsection{Test Suites Quality Assurance}
To assess the quality of the test cases, we randomly sampled 50 instances and two authors independently evaluate whether each test faithfully captures the concern expressed in the corresponding human review comment.
For each instance, annotators were provided with the generated test, the original review comment, and the relevant diff hunk, and were allowed to check the original pull request when additional context was necessary. 
The annotators achieved an agreement rate of 84\%, indicating that the generated tests generally align well with human-identified issues. 
The remaining disagreement cases arise from partial coverage of the review comment and occasional overfitting of tests to specific implementations.

\section{Results}
\label{sec:results}

We aim to answer the following research questions:
\begin{itemize}[leftmargin=12pt]
    \item {\em RQ1: How do different automated review tools perform on \name?} This RQ quantitatively evaluates how effectively automated review tools can capture issues raised by humans, as measured by their ability to help a coding agent pass tests derived from human reviews.
    \item {\em RQ2: How does reviews generated from the automated review tools differ compared to human reviews?} The goal in this RQ is to analyze the types of reviews raised by automated tools and humans and analyze the differences between them. 
\end{itemize}

\subsection{Experimental Settings}

\subsubsection{Evaluated Tools}
We select four representative automated code review tools spanning both open-source and proprietary solutions that are actively used in practice: {Claude Code}\footnote{\url{https://code.claude.com/docs/en/overview}}, {Codex}\footnote{\url{https://openai.com/codex}}, {Devin Review}\footnote{\url{https://app.devin.ai/review}}, and {PR-Agent}\footnote{\url{https://github.com/qodo-ai/pr-agent}}. PR-Agent is an open-source tool, while the remaining tools are proprietary solutions.
We run each review tool to produce per-instance review comments on the 184 PR instances in \name.



\subsubsection{Evaluation Configuration and Environment}

We keep the coding agent used in benchmark curation validation (Section~\ref{sec:curation-validation}) and review evaluation (Section~\ref{sec:eval-workflow}) consistent. 
Both steps use Claude Code with a fixed Sonnet-4.6 backend.
All experiments were conducted on a Ubuntu 20.04 server.

\subsection{RQ1: Evaluation Scores on \name}
\label{sec:rq1}
Table~\ref{tab:RQ1} presents the performance of automated code review tools on \name. Recall that each test encodes a concern originally raised by a human reviewer. Therefore, a higher pass rate indicates that the review comments produced by a tool more often guide the coding agent toward changes that address concerns similar to those identified by human reviewers.

\begin{table}[t]
\centering
\small
\caption{Performance comparison of automated review tools.}
\begin{tabular}{l rr ccc}
\toprule
 & \multicolumn{2}{c}{\textbf{\# Comments}} & \multicolumn{3}{c}{\textbf{Pass Rate (\%)}} \\
Reviewer & Total & Avg/PR  & Behavioral & Structural & Overall\\
\midrule
\textit{Automated Tool} \\
\quad Claude-Code & 1336 & 7.3 & 38.1 & 30.7 & 32.1 \\
\quad Codex & 324 & 1.8  & 38.1 & 16.1 & 20.1 \\
\quad Devin & 1344 & 7.3 & 31.0 & 23.4 & 24.8 \\
\quad PR-Agent & 524 & 2.8 & 38.1 & 19.8 & 23.1 \\
\midrule
Human & 234 & 1.3 & 100 & 100 & 100 \\
\bottomrule
\end{tabular}
\label{tab:RQ1}
\end{table}

\begin{figure}[t]
    \centering
    \includegraphics[width=0.7\linewidth]{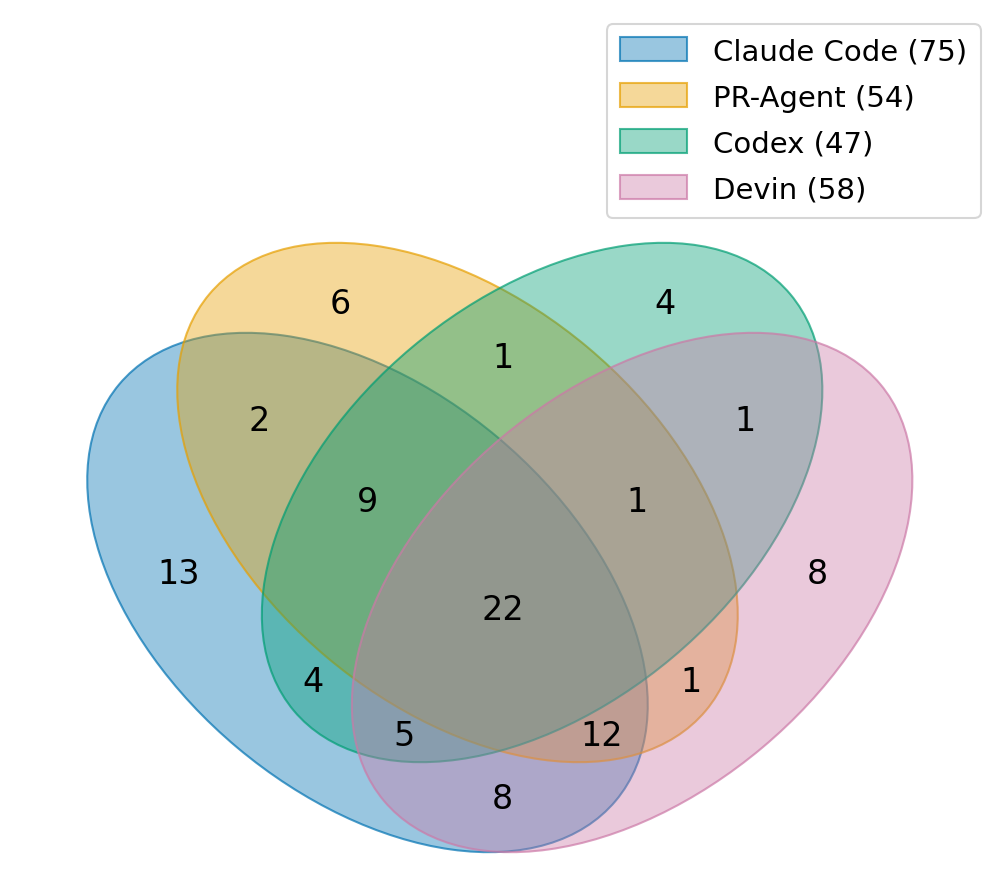}
    \caption{Overlap of passed test for each review tool.}
    \label{fig:pass-rate-overlap}
\end{figure}

\noindent{-\quad {\em Clear gap between human and automated review tools.} }
The results show that the automated review tools achieve pass rates ranging from 20.1\% to 32.1\%, whereas human reviewers achieve a pass rate of 100\%. 
Figure~\ref{fig:pass-rate-overlap} shows the overlap of the passed tests across the four review tools.
Considering the union across all four tools, 97 out of the 234 tests were passed by at least one tool (41.5\%), which shows there is still a substantial gap between AI-generated reviews and human reviews, even when multiple tools are combined. 

\noindent{-\quad {\em Interpreting the gap: automated tools do not raise the same concerns as humans.} }This gap should be interpreted carefully. The lower pass rates do not necessarily imply that the generated reviews are low quality and not useful. Instead, they suggest that the reviews produced by automated tools are less aligned with the types of issues that human reviewers typically raise. This difference indicates that automated review tools and humans may be complementary, opening opportunities to combine both within the software development life cycle. We investigate this difference further in RQ2.

\begin{table}[t]
\centering
\small
\caption{Manually labelled usefulness of review comments generated by four AI code review tools across six PRs.}
\label{tab:validity}
\begin{tabular}{l r r r}
\toprule
\textbf{Tool} & \textbf{\# Comments} & \textbf{\# Useful} & \textbf{Useful \%} \\
\midrule
Claude Code & 40 & 31 & 78\% \\
Codex       &  8 &  7 & 88\% \\
Devin       & 26 & 22 & 85\% \\
PR-Agent    & 18 & 17 & 94\% \\
\midrule
\textbf{Total} & 92 & 77 & 84\% \\
\bottomrule
\end{tabular}
\end{table}

\noindent{-\quad {\em Reviews from automated review tools are useful.}} To better interpret the low pass rates, we manually inspected a sample of 92 review comments from the evaluated tools on six randomly selected PRs. 
Two annotators independently examined each PR along with its generated review comments, with access to the full repository. 
They then labeled each comment as useful or not, where useful comments are those that point out valid issues or suggest improvements to the code.
Disagreements were resolved through discussion. The result in Table~\ref{tab:validity} shows that most comments are useful. This finding indicates that the low pass rates are not a result of poor-quality reviews, but rather because automated review tools tend to highlight different concerns during the review process compared to human reviewers. We investigate this difference further in RQ2.

\rqans{Answer to RQ1:}{There exists a clear gap between reviews generated by automated tools and reviews produced by human.}

\subsection{RQ2: Categories of Review Comments}

\begin{figure*}[h]
    \centering
    \includegraphics[width=0.85\linewidth]{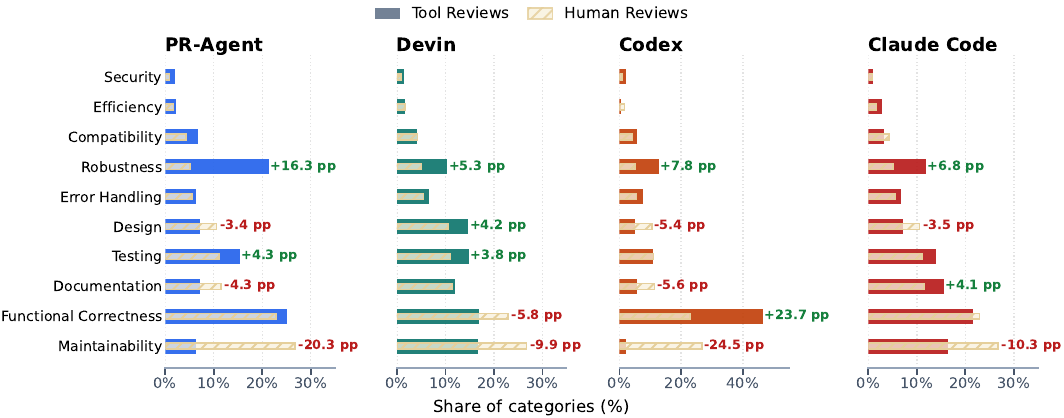}
    \caption{Categories of review comments generated by four tools and comparison with human reviews.}
    \label{fig:rq2}
\end{figure*}

\begin{figure*}[h]
    \centering
    \includegraphics[width=0.85\linewidth]{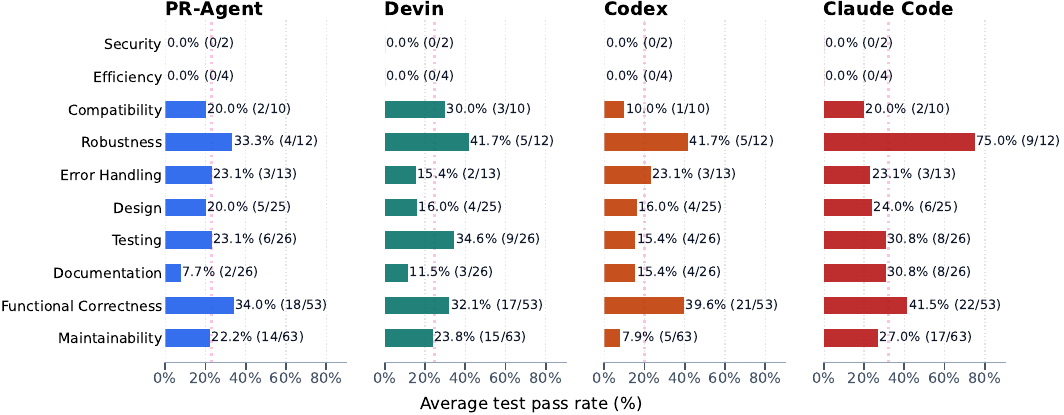}
    \caption{Pass rate of automated review tools grouped by the category of the executable tests.}
    \label{fig:rq2_2}
\end{figure*}

For RQ2, we aim to analyze the alignment between human and automated code reviews by assigning a category label to each review comment. The categorization is done manually with the following two-staged protocol. In stage 1, two authors individually perform preliminary inspections on human reviews in the base SWE-CARE dataset, and arrived at the following categories.
\begin{enumerate}[leftmargin=*]
    \item \textbf{Functional Correctness:} Concerns about incorrect or unexpected high-level behavior, regressions, semantic mismatches, or functional bugs.
    \item \textbf{Testing:} Concerns about incorrect, missing, fragile or flaky test functions, weak assertions, lack of test coverage, or requests improvements of test functions or files.
    \item \textbf{Robustness:} Concerns about implementation-level edge cases, such as handling of corner cases in input and control flow.
    \item \textbf{Compatibility:} Concerns about version, software or hardware compatibility.
    \item \textbf{Documentation:} Concerns about correctness, wording, structure, or clarity in docs, examples, links, or code comments.
    \item \textbf{Design:} Concerns about interaction with other parts of the software, such as API interface signature and return values.
    \item \textbf{Error Handling:} Concerns about warning/error messages or exception behavior.
    \item \textbf{Maintainability:} Concerns about readability, information logging, code smells, code complexity, code organization, and consistency with repository-specific coding practices.
    \item \textbf{Efficiency:} Concerns about overhead and computational efficiency.
    \item \textbf{Security:} Concerns about unsafe behavior, security exposure, or safety issues.

\end{enumerate}
In stage 2, the same authors individually assign each review comment in the dataset to one of the categories defined in stage 1. Then they discuss and resolve the differences.

\smallskip

\noindent{-\quad {\em The distribution of review categories produced by automated tools differs from that of human reviewers.} }Figure~\ref{fig:rq2} presents the percentage distribution of review comment categories from automated code review tools. Automated tools produce proportions of comments on \textit{security}, \textit{efficiency}, \textit{compatibility}, and \textit{error handling} that are comparable to those of humans; however, they address \textit{design}, \textit{documentation}, and \textit{maintainability} substantially less frequently.
A primary reason for this under representation is that these aspects are often highly specific to individual code repositories and reflect coding practices accumulated over time. This observation highlights a potential gap in automated tools’ understanding of repository-level coding conventions. From another perspective, automated tools tend to raise concerns related to \textit{robustness} and \textit{testing} more frequently than humans, suggesting a greater sensitivity to potential corner cases. While such complementary suggestions can be beneficial, the resulting false positives may also substantially increase the burden on maintainers during software development.

\noindent{-\quad {\em Automated tools capture human-identified issues with varying efficacy across categories.}} 
While previous analysis examins \emph{what} types of issues automated tools tend to raise, we now investigate whether these tools successfully capture the issues that human reviewers identified per category. To this end, Figure~\ref{fig:rq2_2} reports the pass rate of automated review tools when benchmark instances are grouped by the category of the corresponding human review comments. 
Recall that each benchmark instance is associated with an executable test derived from a human review. Since human reviews are categorized during the previous labeling process, each executable test inherits the same category. Therefore, the pass rate for each category reflects how often automated reviews successfully capture issues identified by human reviewers.
The results show that automated tools do not capture human concerns uniformly across categories. Overall, across tools, high pass rate is observed on robustness and functional correctness. One possible explanation is that these categories are more localized in the patch under review and this makes them easier to capture. 
In contrast, low pass rate is observed on documentation and design.
The pass rate on maintainability is also limited, ranging from 7.9\% to 27.0\%. These categories often require on broader repository context and project-specific conventions, which current repositories do not document this well and current tools do not utilize this well.

\rqans{Answer to RQ2:}{Automated review tools differ from human reviewers both in the distribution of issue categories they raise and in the categories of human concerns they capture.}

\section{Actionable Insights}
\label{sec:discussion}

\noindent 
{\em a. Automated review tools should be viewed as complements to human reviewers rather than replacements.} 
In RQ1, all evaluated tools achieve substantially lower pass rates than human reviewers, despite often producing more review comments. Further analysis in RQ2 suggests that this gap is at least partly explained by differences in the types of issues raised: automated tools tend to emphasize certain categories (e.g., robustness and testing) while does not cover well others that humans frequently identify. Taken together, these findings point to the possibility for division of labor in future review workflows.

\noindent
{\em b. Repository-specific context is a likely missing ingredient for improving automated review tools.}
In RQ2, automated tools rarely raise issues related to maintainability, design, and documentation, categories that often depend on knowledge specific to a repository. A concrete example can be found in a human review from the Lightning-AI project\footnote{\url{https://github.com/Lightning-AI/pytorch-lightning/pull/4654\#discussion_r526484745}}, where the reviewer suggested using \texttt{os.getenv('COLAB\_GPU')} instead of \texttt{IS\_COLAB} to maintain consistency with existing code. This is not a generic correctness issue; rather, it reflects a local coding convention that is obvious to maintainers but difficult for an automated reviewer to infer from the patch alone. This observation has two implications. For practitioners, documenting repository-specific rules in accessible artifacts, such as architecture notes, style guides, or agent-facing instructions (e.g., \texttt{AGENTS.md}~\cite{DBLP:journals/corr/abs-2601-20404}), may improve the effectiveness of automated review tools. For tool developers, review agents should be grounded in richer project context, including architecture documents, prior review history, naming conventions, and similar code patterns, so that generated feedback can better align with human expectations.

\noindent
{\em c. Tests can be useful artifacts for building code review agents.}
A key advantage of our benchmark is that it evaluates review comments based on whether they lead to code changes that resolve issues identified during the review process, as captured by executable tests. This provides a clear and objective optimization target for evaluation, allowing automated review tools to be assessed using execution-based signals rather than relying solely on textual similarity or subjective LLM-as-a-judge assessments. More broadly, executable tests also suggest a useful future direction for training review systems, since they reward issue identification that leads to verifiable improvements in code. 

\section{Threats to Validity}

\noindent
{\em a. Internal Validity.} 
Our evaluation relies on a coding agent to revise patches based on review comments. Therefore, the outcomes may reflect not only the quality of the reviews but also the capability of the agent. 
To mitigate this, we validate benchmark instances using human-written comments prior to inclusion, and the same coding agent is used in both the validation and the evaluation pipeline. 
This ensures that all retained instances are solvable.

\noindent
{\em b. External Validity.}
Our benchmark is constructed from SWE-CARE and includes 184 pull request instances with 234 verifiable oracles across 56 repositories. While this curation improves reliability, it may limit representativeness with respect to the broader software ecosystem. Nevertheless, the dataset can be easily extended, as the pipeline can be applied to new pull request instances.

\noindent
{\em c. Construct Validity.}
The filtering stage in our pipeline leverages an LLM-based classifier, which may retain borderline comments. To mitigate this, we leverage a strict fail-then-pass criterion during test generation and verify whether each problem can be solved during the validation phase using the original human comments. Moreover, manual analysis of the reviews involve human judgment. To mitigate this, two authors independently performed labeling.

\section{Conclusion}
\label{sec:conclusion}
The rapid advancement of AI-based code generation has made generating code becomes increasingly easy and scalable. On the other hand, the process of reviewing the code remains a bottleneck. Several automated code review tools have been proposed, but evaluating their effectiveness is an open challenge.

In this paper, we revisit the problem of evaluating code review tools. We propose \name, a benchmark designed to assess how well automated review tools identify the same issues that human reviewers raise in realistic PR settings. \name adopts a test-based evaluation, where it converts human review comments into executable tests. Our empirical results show that current state-of-the-art review tools identify only a fraction of the issues captured in the benchmark. This finding highlights the limitations in existing code review tools. Further analysis reveals that tool-generated reviews often highlight different aspects of code compared to human reviewers. This suggests that human and automated review tools are complementary to each other. Looking forward, our findings suggest several directions for future research, such as integrating automated tools into human-centric review workflows and improving the ability of code review tools to better capture relevant context in the codebase.

\section*{Data Availability}
\label{sec:data-availability}
Our replication package, including source code, benchmark datasets and results, is available at \arxivurl.

\begin{acks}
This work was partially supported by a Singapore Ministry of Education (MoE) Tier 3 grant “Automated Program Repair”, MOE-MOET32021-0001.
\end{acks}

\section*{Disclaimer}
The views and conclusions expressed in this paper are those of the authors alone and do not represent the official policies or endorsements of SonarSource. 
Furthermore, the findings presented herein are independent and should not be interpreted as an evaluation of the quality of products at SonarSource.

\bibliographystyle{ACM-Reference-Format}
\bibliography{ref}
\end{document}